\documentclass[aps,12pt,nofootinbib]{revtex4}
\usepackage{graphics}
\usepackage[dvips]{color}
\usepackage{amsmath}
\usepackage{amssymb}

\usepackage{epsfig}

\def\beq{\begin{equation}}
\def\eeq{\end{equation}}

\def\bea{\arraycolsep .1em \begin{eqnarray}}
\def\eea{\end{eqnarray}}

\newcommand{\step}{\vspace{.5em}}

\def\s0#1#2{\mbox{\small{$ \frac{#1}{#2} $}}}
\def\0#1#2{\frac{#1}{#2}}

\def\grgl{\:\hbox to -0.2pt{\lower2.5pt\hbox{$\sim$}\hss}{\raise3pt\hbox{$>$}}\:}
\def\klgl{\:\hbox to -0.2pt{\lower2.5pt\hbox{$\sim$}\hss}{\raise3pt\hbox{$<$}}\:}

\newcommand \be {\begin{equation}}
\newcommand \ee {\end{equation}}
\newcommand \bed {\begin{displaymath}}
\newcommand \eed {\end{displaymath}}

\newcommand{\bit}{\begin{itemize}}
\newcommand{\eit}{\end{itemize}}


\begin{document}

\title{ Enhancement of field renormalization in scalar theories via  functional
renormalization group}

\author{Dario Zappal\`a}

\affiliation{
\mbox{INFN, Sezione di Catania, 64 via S. Sofia, I-95123, Catania, Italy.}
}%

\begin{abstract}
${}$\\[-1ex]
\centerline{\bf Abstract\hskip 90 pt}
The flow equations of the Functional Renormalization Group are applied to
the $O(N)$-symmetric scalar theory, for $N=1$ and $N=4$, in four Euclidean dimensions,
$d=4$,
to determine the effective potential and the renormalization function of the field in the
broken phase.
In our numerical analysis, the infrared limit, corresponding to the vanishing of the
running momentum scale in the equations, is approached to obtain the physical
values of the parameters by extrapolation.
In the $N=4$ theory a non-perturbatively large value of the physical
renormalization of the longitudinal component of the field is observed.
The dependence of the field renormalization on the UV cut-off and on the
bare coupling is also investigated.
\end{abstract}

\pagestyle{plain} \setcounter{page}{1}

\maketitle

\section{Introduction}

An important technique in continuum field theory, is the functional
renormalization group  \cite{Polchinski:1983gv,Wetterich:1992yh},
which represents a powerful method to approach both perturbative and
non-perturbative phenomena; it is based  on the infinitesimal
integration of momentum modes from a path integral representation of
the theory with the help of a Wilsonian momentum cutoff
\cite{Wilson:1973jj}. The resulting functional flow equations
interpolate between the  microscopic theory at short distances and
the full quantum effective theory at large distances. \step

In the past years various realizations of the functional renormalization
group were developed and, among the most renowned, there
are Polchinski's and  Wetterich's  formulations
(for reviews see \cite{Bagnuls:2000ae} \cite{Berges:2000ew}).
At the same time, the derivative expansion,
a powerful  approximation scheme that relies  on a small anomalous dimension of the
field,  was introduced to reduce the full flow equation to a small set of
treatable partial differential equations \cite{Morris:1994ie,Litim:2001dt},
More recently, a different approximation scheme was presented in
\cite{blaizot,Benitez:2009xg,Benitez:2011xx}, where the flow, rather than being
projected on the semi-local derivative expansion of the action,
is projected on the $n$-point 1-particle irreducible (1PI) vertices,
so that  the momentum  dependence of the correlation functions is properly taken into
account.

\step

The range of application of the functional renormalization group is wide
and it is especially suitable for the study of phase transitions, due to its
flexibility  even in the presence of strong correlations or couplings.
For instance fixed point studies of Ising-like or $O(N)$  theories, including the
determination of universal critical indices, were repeatedly carried out,
using different versions of the renormalization group (RG) flow equation
(see for instance \cite{Litim:2010tt} and references therein),
both to analyze the properties of the theory and to test the accuracy of the method
through a comparison with other field-theoretical techniques (see e.g. 
\cite{zinn,Pelissetto,ParisenToldin:2003hq}).

\step

Some attention has also been devoted to
the study of the non-perturbative features of the effective action and effective
potential in the presence of spontaneous symmetry breaking.
In fact is it a well known result of quantum field
theory \cite{syma,ilio,curt} that the  effective potential defined through a Legendre
transformation must be convex, a property that, in the limit of infinite volume,
leads to a non-analytic behavior with a typical degenerate flat region of the
potential in correspondence of the classically forbidden region between the
classical minima. The convexity of the potential  can only be recovered by resorting
to some non-perturbative scheme (see e.g. \cite{convexity})
and it has already been studied  by means of RG techniques
\cite{ringwald,tetradis1,aoki,alexander,tetradis2,d1,bonannolac,
Litim:2006nn,consolizap,caillol}.
\step

In this paper we analyze in detail the spontaneously broken phase of the
$O(N)$-symmetric scalar theory, for $N=1$ and $N=4$,  in four Euclidean dimensions, $d=4$,
going beyond the Local Potential
Approximation (LPA) and including the effects of the field renormalization.
Instead of deriving the flow equations for the potential $V$ and the
field renormalization $Z$ directly from an expansion of the full flow equation
for the effective action around a constant field configuration, we shall derive them
starting from the approach developed in
\cite{blaizot,Benitez:2009xg,Benitez:2011xx}
and adopting the regulator for the infrared modes
introduced in
\cite{Litim:2000ci,Litim:2001up,Litim:2001fd,Pawlowski:2005xe},
which has the specific property of optimizing the determination of
the critical exponents
in the LPA. This particular choice allows us to analytically perform
the integration over the momentum variable,
which simplifies the structure of coupled Partial Differential Equations (PDEs)
for $V$ and $Z$.
The aim of the present analysis is to provide a better understanding
of the details of this phenomenon which, for $N=4$ and $d=4$, could
provide some indications on the description of the broken chiral
symmetry phase for a two-flavor quark model or on  the mass
generation mechanism in the Standard Model, although in both cases
the similarity has serious limitations. In fact, in the former case
the $O(4)$ model is in the correct universality class to describe
the universal features of the chiral phase transition, but when the 
system is not close to the critical line, which is the case 
considered here, then some properties of the fermionic model could be
poorly reproduced by the scalar fields. Concerning the latter case,
the coupling to gauge fields is turned off in the $O(N)$ model and
therefore the massless Golstone bosons play an important role in the
infrared sector of the ordered phase and this could lead to a different
behavior with respect to the usual Higgs mechanism where the
Goldstone are replaced by the longitudinal degrees of freedom of the
massive gauge bosons which, due to their mass, are less relevant for
the infrared dynamics.

\step

An attempt to use the two coupled  flow equations for $V$ and $Z$
in this context was already done in \cite{d1} for the simple quanto-mechanical
problem ($d=0+1$ dimensions) and in \cite{consolizap} for the $N=1$ scalar theory in
$d=4$ dimensions. In both cases the Proper Time version of the RG flow
\cite{Liao:1996fp,Bohr:2000gp,Bonanno:2000yp,Litim:2001hk,Mazza:2001bp,Bonanno:2004sy}
(which can be regarded as an approximation to the class of exact background field
flows \cite{Litim:2001ky,Litim:2002xm,Litim:2002hj,Litim:2006nn}) was adopted
in the exponential form, which is derived by using a sharp cut-off on the
Proper Time variable \cite{Litim:2001hk,d2}.
In the quanto-mechanical case the corrections induced by the inclusion of
$Z$ turned out to be in  very good agreement with the (exact) Schroedinger equation
output  while, for the quantum field theory, a very large $Z$ was observed in
the classically forbidden region. However, in the latter case, only partial conclusions
could be drawn, because the numerical  instabilities in the flow equations made
rather difficult the approach to the infrared (IR) limit,  where all quantum modes
are integrated out and the full field renormalization is obtained.
On the contrary, the approach used
in this paper, based on a different approximation scheme, turns out to be numerically
more stable, thus allowing us to push the RG scale low enough to 
get a better insight of the IR region.

\step

The scheme of the paper is the following. In Sect. II the main
details of the RG flow equations and of the approximation used are
recalled; the results of the numerical analysis of the potential and
the field renormalization for the single field,  $N=1$, and for the
$N=4$ theory in $d=4$ are shown in Sect. III. The conclusions are
reported in Sect. IV. 

\step

\section{RG flow equations}
\label{RGFLOW}

The starting point of our analysis
 is the partition function of the theory, defined as a functional
of the source $J(x)$, and also dependent on the momentum scale $k$:
\beq\label{zetacappa}
{\cal Z}_k[J]=\int {\cal D}\varphi \, {\rm e}^{-S-\Delta S_k+\int_x J \varphi}\, .
\eeq
where the usual action $S$ is modified by the $k$ dependent
regulator, quadratic in the field $\varphi$, $\Delta S_k[\varphi]$:
\beq\Delta S_k[\varphi]= \frac{1}{2} \int_q\: R_k(q)\varphi(q)\varphi(-q), \eeq
with
\beq \int_q\equiv\int \frac{d^dq}{(2\pi)^d} \;\;\;\;\;\;
{\rm and }\;\;\;\;\;\; \int_x\equiv\int {d^dx}\;.
\eeq

In order to obtain a physically relevant flow, $R_k(q)$ must suppress the
modes with $q\ll  k$ while allowing to integrate those with  $q\gg k$
and therefore one can choose  $R_k(q)$ of order $k^2$ in the former region
and $R_k(q)\sim 0$ in the latter. In particular at $k=0$ the regulator must
vanish so that  ${\cal Z}_{k=0}[J]$  coincides with the standard
partition function ${\cal Z}[J]$. 
In this framework the following flow equation is obtained \cite{Berges:2000ew}
($ \partial_t\equiv k \partial_k$):
\beq \label{rgfloweq}
\partial_t \Gamma_k[\phi]=\frac{1}{2} \int_q\, \partial_t R_k(q)
\left [ \Gamma_k^{(2)}[q,-q;\phi]+R_k(q)\right ]^{-1}
\eeq
where $\Gamma_k[\phi]$ is the modified effective action at scale $k$,
defined by:
\beq
\int_x J \phi -\Delta S_k[\phi] -\Gamma_k[\phi]    = \log {\cal Z}_k[J]
\eeq
with $\phi(x)=\delta \ln {\cal Z}_k[J]/\delta J(x)$ and
and  $\Gamma_k^{(2)}[q,-q;\phi]$ is the Fourier transform
of  the second functional derivative of $\Gamma_k[\phi]$:
\beq
\label{gamma2}
\Gamma_k^{(2)}[x_1,x_2;\phi]\equiv  \frac{\delta^2\Gamma_k}{\delta\phi(x_1)
\delta\phi(x_2)} \;.
\eeq
Note that the term
\beq
\label{propagator}
G_k(q,\phi)=\left [ \Gamma_k^{(2)}[q,-q;\phi]+R_k(q) \right ]^{-1}
\eeq
appearing in Eq. (\ref{rgfloweq}) is the
full propagator of the action modified by the regulator $\Delta S_k$.
Then one can take the initial condition for the flow equation Eq. (\ref{rgfloweq})
at the scale $k=\Lambda$ where fluctuations are frozen by
$\Delta S_k$, so that $\Gamma_{k=\Lambda}[\phi]\approx S[\phi]$ and
the full effective action $\Gamma[\phi]$ of the original theory is obtained
as the solution of  Eq. (\ref{rgfloweq}) when  $k\to 0$, where $R_k(q)$
vanishes and all fluctuations have been integrated out.

\step

The most straightforward approximation scheme to treat
the RG flow is to express the effective action $\Gamma_k[\phi]$
through an expansion  in the derivatives of the field:
\begin{equation}
\Gamma_k[\phi]=\int_x\left(  V_k(\phi)  +\frac{1}{2}
Z_k(\phi)\left( \partial \phi \right)^2 + O(\partial^4) \right).
\label{dexpansion}
\end{equation}
and to project the full equation (\ref{rgfloweq}) onto a set of flow  equations
for the coefficients of the expansion $V_k(\phi)$, $Z_k(\phi)$, ...\; .
This projection requires an expansion of the right hand side of
Eq. (\ref{rgfloweq}) in powers of the field derivatives or, equivalently,
in powers of the momentum.

\step

An alternative scheme has been developed in
\cite{blaizot,Benitez:2009xg,Benitez:2011xx}
and, below, we shall briefly recall its essential features
(in the rest of this Section we will follow the notation
adopted in \cite{Benitez:2011xx} ). This scheme consists in
an expansion in terms of the $n$-point functions, $\Gamma^{(n)}_k$,
i.e. the $n$-th functional
derivative of $\Gamma_k[\phi]$ with respect to the field $\phi$.
For instance the first two equations for $n=0$ and $n=2$, correspond
to the two following equations, respectively for the potential $V_k$ defined in
Eq. (\ref{dexpansion}) and equal, up to a volume  factor, to $\Gamma_k$
evaluated at a constant field configuration $\phi$, and the 2-point function
computed at constant $\phi$ :

\beq\label{vflow}
\partial_t V_k(\phi)=\frac{1}{2}\int _q \,\partial_t R_k(q)\, G_k(q,\phi),
\eeq
\begin{widetext}
\begin{eqnarray}
\label{gamma2flow}
\partial_t\Gamma_{k}^{(2)}(p,\phi)=\int_q
\partial_t R_k(q) G_{k}(q,\phi)\bigg\{&&\hspace{-4mm}
\Gamma_{k}^{(3)}(p,q,-p-q,\phi) G_{k}(q+p,\phi)
\Gamma_{k}^{(3)}(-p,p+q,-q,\phi)\nonumber\\
&&\hspace{-4mm} -\frac{1}{2}\Gamma_{k}^{(4)}
(p,-p,q,-q,\phi)\bigg\} G_{k}(q,\phi)\; .
\end{eqnarray}
\end{widetext}

It is easy to check from the functional structure of Eq. (\ref{rgfloweq})
that the flow equation of $\Gamma^{(n)}_k$ involves the $s$-point functions
with $s\leq n+2$.
Therefore an approximation or truncation is needed
to obtain a closed set of PDEs. As explained in
\cite{blaizot,Benitez:2009xg,Benitez:2011xx} this can be realized at any fixed $n$,
by neglecting the internal momentum $q$ in $\Gamma^{(n+1)}_k$ and  $\Gamma^{(n+2)}_k$,
so that they can be written in terms of $\Gamma^{(n)}_k$ according to
the following property, valid for constant field configurations :
\beq\label{relazione}
\Gamma^{(n+1)}_k(\left\{p_i\right\},0,\phi)=
\partial_\phi \Gamma^{(n)}_k(\left\{p_i\right\},\phi)
\eeq
where the index $i$ runs between 1 and $n$.
In particular, for $n=2$, $\Gamma^{(3)}_k$ and  $\Gamma^{(4)}_k$
in Eq. (\ref{gamma2flow}) are computed at $q=0$ and expressed in terms of
$\Gamma^{(2)}_k$, according to Eq. (\ref{relazione}). Thus, a closed set of two
equations for $ V_k$ and  $\Gamma^{(2)}_k$ is explicitly obtained in 
\cite{blaizot,Benitez:2009xg,Benitez:2011xx}.

\step

Our aim is to extract the coupled flow equations of $ V_k$ and  $Z_k$
from those of  $ V_k$ and  $\Gamma^{(2)}_k$. This is easily obtained
by performing a derivative expansion of the latter, to order $O(p^2)$.
To this purpose we restrict the form of $\Gamma^{(2)}_k$ according to the  ansatze in
Eq. (\ref{dexpansion}), i.e.:
\beq\label{newansatz}
\Gamma^{(2)}_k (p,\phi)= Z_k(\phi)  p^2 + V_k^{\prime\prime}(\phi) + O(p^4)
\eeq
where the prime indicates the derivative with respect to the field $\phi$. Then,
the flow equation of $V_k$ is given in  Eq. (\ref{vflow}),
while the insertion of Eq. (\ref{newansatz}) into Eq. (\ref{gamma2flow})  gives
\bea
\label{zetaflow}
&p^2 \partial_t\ Z_k(\phi)+ O(p^4)=
J_3(p,\phi) \; \left[V_k^{\prime\prime\prime}(\phi)+p^2 Z_k^\prime(p,\phi)\right]^2-
\nonumber\\
&I_3(\phi)\; \left ( V_k^{\prime\prime\prime}(\phi) \right )^2
-\frac{1}{2} I_2(\phi) \; p^2 Z_k^{\prime\prime}(p,\phi)+ O(p^4)\;,
\eea
where we have used the notation of \cite{Benitez:2011xx} :
\beq\label{integrali}
I_n(\phi) \equiv J_n(p=0,\phi) \;\;\;\;{\rm and }  \;\;\;\;
J_n(p,\phi )\equiv \int_q \;
\partial_t R_k(q)\; G_k(p+q,\phi )G^{n-1}_k(q,\phi)
\eeq
with $G_k$  given in Eq. (\ref{propagator}) and
$\Gamma^{(2)}_k$  in Eq. (\ref{newansatz}).
Finally, to obtain the flow equation of $Z_k$
one has to expand the integral $J_3(p,\phi)$
in powers of $p^2$ and systematically neglect the $O(p^4)$ terms
in  Eq. (\ref{zetaflow}) and, as the terms proportional to $p^0$ in the right
hand side vanish, one reads the equation for $Z_k(\phi)$  from the
$O(p^2)$ terms.
Then, the PDEs coming from 
the derivative
expansion to order $O(p^2)$ of the 
$\Gamma^{(2)}_k$ equation derived in
\cite{blaizot,Benitez:2009xg,Benitez:2011xx}, contain the same
approximation made on the latter, namely the neglect of the internal
momentum $q$ dependence of  $\Gamma^{(3)}_k$, $\Gamma^{(4)}_k$.

\step

The extension to the $O(N)$-symmetric scalar theory is straightforward
and it is also illustrated in \cite{Benitez:2011xx}.
According to the symmetry of the theory the propagator can be written
in terms of its longitudinal (L) and transverse (T) components
with respect to the external field
\beq\label{propagatoren}
G_{ab}(p^2,\phi)=G_T(p^2,\rho)\left(\delta_{ab}-\frac{\phi_a\phi_b}{2\rho}\right)
+ G_L(p^2,\rho)\frac{\phi_a\phi_b}{2\rho}.
\eeq
where
\beq
\rho\equiv\frac 1 2 \phi_a \phi_a
\eeq

If the 2-point function is parametrized as
\beq\label{gamma2n}
\Gamma_{ab}^{(2)} (p, -p,\rho) = \Gamma_{A} (p,\rho)\delta_{ab} +
\phi_a \phi_b \Gamma_B (p, \rho)
\eeq
and $\Gamma_A$ and $\Gamma_B$ are expressed in the following way
(here the dots indicate derivatives with respect to $\rho$ and the script $k$
is omitted for simplicity)
\beq
\Gamma_A(p, \rho) =  Z_A( \rho) p^2 +  \dot V  \label{gammaab}
\;\;\;\;\;\;\; {\rm and } \;\;\;\;\;\;\;
\Gamma_B(p, \rho) = Z_B (\rho) p^2 + \ddot V \; ,
\eeq
then the following relations hold
\beq\label{gt}
G_T^{-1}(p,\rho) = \Gamma_A(p,\rho) + R_k(p) = Z_T( \rho) p^2 + \dot V
+ R_k(p)
\eeq
\beq\label{gl}
G_L^{-1}(p,\rho) = \Gamma_A(p,\rho) + 2\rho \Gamma_B(p,\rho)+ R_k(p)=
Z_L( \rho) p^2 + \dot V  + 2\rho \ddot V+ R_k(p)
\eeq
with $Z_T$ and $Z_L$ defined as
\beq
Z_T = Z_A  \;\;\;\;\;\;\; {\rm and } \;\;\;\;\;\;\; Z_L = Z_A + 2 \rho Z_B \;\;.
\label{zlzt}
\eeq

The  flow equations for $V$, $Z_A$ and $Z_B$  are

\begin{equation}
\label{o4vflow}
\partial_t V(\rho) = \frac{1}{2}\left \{ (N-1)I_1^{TT}(\rho)+I_1^{LL}(\rho) \right \}
\end{equation}

\bea
\label{zaflow}
&p^2 \partial_t Z_A(\rho) + O(p^4) = 2  \rho \Big\{ J_3^{LT} (p^2 \dot Z_A +
\ddot V)^2 + J_3^{TL} (p^2 Z_B+ \ddot V)^2 - (I_3^{LT} +I_{3}^{TL}) \ddot V^2 \Big\}
\nonumber\\
&-\frac{1}{2} I_2^{LL} p^2 (\dot Z_A + 2 \rho \ddot Z_A) -
\frac{1}{2} I_2^{TT} p^2 \left ( (N-1) \dot Z_A + 2 Z_B\right )   + O(p^4) \;,
\eea
\bea\label{zbflow}
&p^2 \partial_t Z_B(\rho) + O(p^4) = J_3^{TT} (N-1) (p^2 Z_B + \ddot V)^2
- J_3^{LT} (p^2 \dot Z_A + \ddot V)^2 - J_3^{TL} (p^2 Z_B+ \ddot V)^2
\nonumber \\
&+ J_3^{LL} \Big\{(p^2 \dot Z_A +2 p^2 Z_B+ 3\ddot V)^2
+ 4 \rho \big(p^2 \dot Z_B + \partial_\rho \ddot V \big)
\big(p^2\dot Z_A+2p^2 Z_B +3 \ddot V\big)
\nonumber \\
&+ 4 \rho^2 (p^2\dot Z_B + \partial_\rho \ddot V)^2\Big\}
- \frac{1}{2} I_2^{TT} (N-1){p^2} \dot Z_B
- \frac{1}{2} I^{LL}_2 {p^2}(5 \dot Z_B + 2 \rho \ddot Z_B)
\nonumber\\
&- \left((N-1) I^{TT}_3-I^{LT}_3-I^{TL}_3\right) \ddot V^2
- I^{LL}_3 (3\ddot V+2\rho \partial_\rho \ddot V )^2 + p^2 Z_B
\left ( I^{LT}_3+I^{TL}_3 \right ) + O(p^4)
\eea
with  the following  definitions of the integrals ($n>1$ and $\alpha$, $\beta$ 
stand either for $L$ or $T$) :
\beq
\label{intj}
J_n^{\alpha \beta}(p,\rho)=
 \int_q \partial_t R_k(q) G_{\alpha}^{n-1}(q,\rho) G_{\beta}(p+q,\rho)
\;\;\;\;\; {\rm and } \;\;\;\;\;
I_n^{\alpha \beta}(\rho)=J_n^{\alpha \beta}(p=0,\rho),
\eeq

The only undefined ingredient in the above flow equations is the explicit form of the
regulator. There are various functional forms of $R_k(p)$,
tested in many specific problems and
whose properties have been  studied in detail. From a practical point
of view, since we are interested in the numerical resolution of a set of
PDEs,  it is preferable to take a particular
regulator which  allows us to solve analytically the momentum integrals.
The cutoff
\cite{Litim:2000ci,Litim:2001up,Litim:2001fd,Pawlowski:2005xe} :
\beq
\label{optimctf}
R_k(q) =(k^2 -q^2) \;\;\Theta (k^2 -q^2)
\eeq
(where $\Theta$ indicates the Heaviside step function)
makes the resolution of the integrals particularly simple and
therefore  we use this regulator in our analysis.

\step

Regarding this choice, some comments are in order.
In fact the flow equations contain the derivative
of the regulator with respect to the scale $k$
and, as discussed above, the flow of $Z_k$ is obtained 
after an expansion in powers of the external momentum $p$
which, again, introduces derivatives of the regulator.
The regulator in Eq. (\ref{optimctf}) has the form 
$x \Theta (x)$ so that its derivative 
$\partial_t R_k(q)$ generates two terms : 
the first one, $(\partial_t \, x)\, \Theta (x)$, which 
produces relevant contributions and the second one 
$ x \, (\partial_t \, \Theta (x))$ which produces a term
proportional to $ x \,\delta (x)$ that vanishes under 
integration over the momentum $q$, as long as  no
pathologies appear in the integrand. Therefore only 
the first term is to be retained in  the flow equations.

\step

Then, when deriving the  PDE  
for  $Z_k$ from the flow of the 2-point function,
an expansion in the momentum $p$ of the integral $J_3(p,\phi)$,
defined in Eq. (\ref{integrali}),
to order $O(p^2$) is necessary and this generates new terms proportional to 
$\Theta$ and $\delta$-functions. Also, a derivative of delta-function 
of the form $\partial_{p_\nu} (y\,\delta(y) )$ (with $y=k^2-(p+q)^2$ ) is generated, which,
due to the particular form  of the variable $y$,
can be replaced by $\partial_{q_\nu} (y\,\delta(y) )$ and the corresponding 
term can be calculated by means of an integration by parts,
by recalling that the structure  $y\,\delta(y)$ gives zero contribution
even when evaluated at the boundaries of the integral over the momentum variable $q$.
After the integration by parts, by making explicit 
the dependence on the distributions, one finds 
(below, $l_{\mu\nu}(p)$ and $h_{\mu\nu}^{(i)}(p,q)$, $i=1,2,3$ 
generically indicate regular functions appearing in the expansion
and, again, $y=k^2-(p+q)^2$) :

\bea
\label{pderivate2}
&\partial_{p_\nu} \partial_{p_\mu} J_3(p,\phi) = 
l_{\mu\nu}(p)\;+\;  
\int^{k^2}_0  \; d q^2 \;\Big\{ h_{\mu\nu}^{(1)}(p,q) \; 
\big [ \Theta (y) +y \; \delta (y)\big ] \Big\} \nonumber\\
&+\; \int^{k^2}_0 \; d q^2 \;\Big\{  h_{\mu\nu}^{(2)}(p,q) \; 
\big [\Theta (y) +y \; \delta (y)\big ] \,
\big [\Theta (y) +y \; \delta (y)\big ]
\; +\; h_{\mu\nu}^{(3)}(p,q) \; \delta (y) \Big\}
\eea
and Eq. (\ref{pderivate2}), evaluated at $p=0$, contributes to the flow equation of $Z_k$.
The terms in Eq. (\ref{pderivate2}) proportional to $\Theta (y)$ or $\Theta^2 (y)$
give finite contributions while, by regarding the 
delta-function as the limit of a sequence of normal distributions, we discarded
all terms proportional to 
$y\,\delta(y)$ or $ (y\,\delta(y))^2 $ or $y\,\delta(y)\, \Theta (y)$, and calculated 
the last term in Eq. (\ref{pderivate2}), where the argument of the delta-function 
coincides with the upper boundary of the integral, according 
to $\int_0^{x_0} f(x) \delta(x-x_0)=f(x_0) /2$.

\step

In Sect. \ref{numerical} we study  the RG flow equation of a scalar theory in
the broken phase in $d=4$, concentrating on
two specific cases: the $N=1$ and the  $N=4$ theory.
We use the ansatze in Eq. (\ref{dexpansion})
with the following initial condition at the ultraviolet (UV)
scale $k=\Lambda$
\beq
\label{explbare}
\Gamma_{k=\Lambda}(\rho)= \int_x \left \{
\frac{1}{2} \partial_\mu \phi_a \partial_\mu \phi_a -
\frac{1}{2}\phi_a\phi_a + \lambda \left (\phi_a\phi_a\right )^2 \right \}
\eeq
where $a=1$  for $N=1$,  and runs from $1$ to $4$  for $N=4$.
According to
Eq. (\ref{explbare}), at $k=\Lambda$ the field renormalization is $1$
(for the $N=4$ theory $Z_L=Z_T=1$ which, in turn means $Z_A=1$, $Z_B=0$)
and the potential $V_{\Lambda}$ only contains
quadratic and quartic terms in the field, i.e. the only renormalizable
terms in $d=4$. In order to restrict ourselves to the broken phase, the bare mass,
related to the quadratic term of the potential in Eq. (\ref{explbare}),
must be negative. In particular, its explicit value is fixed at $-1$ so that,
in the following, all dimensionful variables are automatically
expressed in units of the bare mass.
It must also be recalled that the phase boundary is a function of
the UV cutoff $\Lambda$ and of the bare coupling $\lambda$, and
both parameters should not be increased
too much to avoid a transition to the disordered phase.

\section{Field renormalization in the broken phase}
\label{numerical}

For practical convenience, in the following numerical analysis,
instead of concentrating on the flow of $Z_k(\phi)$ and $V_k(\phi)$,
we focus on  $Z_k(\phi)$ and $V^\prime_k(\phi)$,
i.e. on the derivative  of the potential  with respect to $\phi$,
which means  that for  the $O(N)$ theory, instead of using $\rho$,
we express all variables in terms of
\beq
\label{onphi}
\phi\equiv\sqrt{\phi_a\phi_a} =\sqrt{2\rho}\; .
\eeq
and, in the following we indicate with $\bar \phi$ the field at the minimum of the
potential in the IR limit $k=0$:
\beq
\label{phibardef}
V^\prime_{k=0}(\bar\phi)=0 \; .
\eeq
The Green functions of the theory are obtained from the derivatives
of the effective action evaluated at the minimum configuration $\phi=\bar\phi$
and therefore $\bar\phi$ is crucial in the  determination
of physical observables.
\step

So, in $N=1$  we solve the two coupled  PDEs given by
Eq. (\ref{zetaflow}) and the field derivative of Eq. (\ref{vflow}),
while for the $N=4$ theory we solve the set of three PDEs
for $Z_A,\,Z_B$ in Eqs. (\ref{zaflow}, \ref{zbflow}) and the field
derivative of Eq. (\ref{o4vflow}) and in all cases  $d=4$.
The numerical solution of our set of PDEs is obtained with the help
of the NAG routine \cite{nag} that integrates a system of non-linear parabolic
partial differential equations in the $(x,t)$ two-dimensional plane. The
spatial discretization is performed using a Chebyshev $C^o$
collocation method, and the method of lines is employed to reduce
the problem to a system of ordinary differential equations. Typically we used 
Chebyshev polynomials of order 3 or 4  which already provide a stable solution.
The routine contains two main parameters - the number of mesh points on the 'space'
axis and the local accuracy $\Delta$ in the 'time' integration - that can be adjusted 
to control the stability of the solution. The first parameter is taken in such a way
that the distance between two subsequent mesh points is  
$\delta x \sim 10^{-3}, \; 10^{-4}$  while $\Delta$  typically is taken  between
$10^{-7}$ and $10^{-9}$. The size of the time step is adjusted at each integration 
step by the routine to keep the accuracy below $\Delta$.
The number of time steps required to converge is strongly dependent on the problem considered, ranging from few hundreds up to $10^{7}$ for the 
hardest cases.

\step 

\begin{figure}
${}$\vskip1cm \epsfig{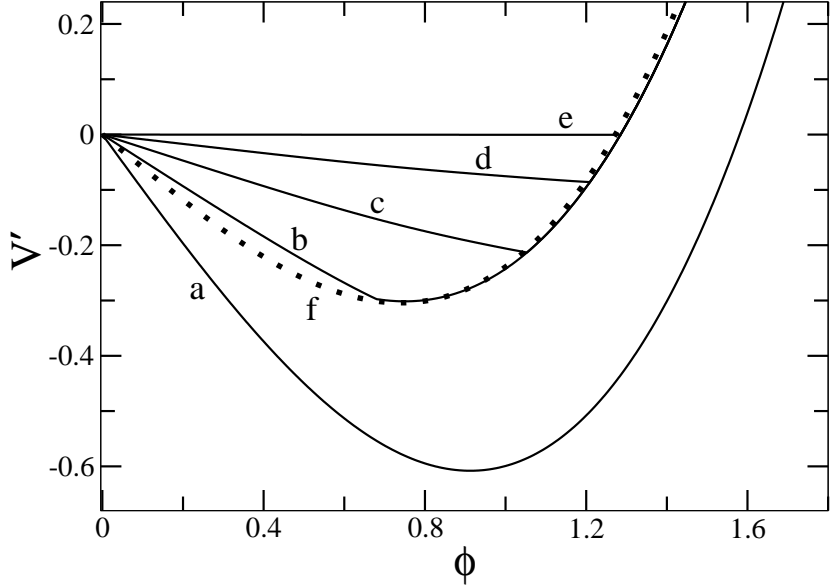}
\caption{\label{uno}
$V^\prime_k(\phi)$ for $N=1$ and $d=4$
at various $k$: $k=\Lambda{\bf (a)},
0.7{\bf (b)},\, 0.5{\bf (c)},\,0.3{\bf (d)},\, 0.02{\bf (e)}$, with $\Lambda=10$
and $\lambda=0.1$. The dotted curve ${\bf (f)}$  is the one loop effective potential derivative,
$V_{1l}^\prime(\phi)$, in terms of renormalized parameters.
\\[2ex]}
\end{figure}

The resolution of the PDEs is done with  the 'space' variable constrained between
$0<\phi<\phi_{bound}$ and for our purposes   $\phi_{bound}=4$ was sufficiently large
to ensure the stability of the solution. The required initial conditions 
for the PDEs at initial 'time' $k=\Lambda$ and $0<\phi<\phi_{bound}$ are 
implicitly given in Eq. (\ref{explbare}). The boundaries at $\phi=0$
in  the entire range of 'time' integration reflect the required symmetry of our model:
$V_k^{\prime}(0)=Z_k^{\prime}(0)=0$ while at $\phi=\phi_{bound}$ we  keep 
free boundaries. 
Finally, the 'time' integration that starts at $k=\Lambda$ should, in principle, reach
$k=0$ or at least should be pushed close enough to zero that the solution has 
become invariant. But in the following analysis, due to the not smooth behavior 
of the solution, which gets worse when approaching the point $k=0$, the numerical integration breaks down well before $k=0$. Clearly this is peculiar of the broken phase 
because in the symmetric phase, where both solution for $V_k$ and $Z_k$ are much
smoother, one can easily reach values of the scale $k$ where the solutions are already
stable.
In the broken phase the lowest accessible value of $k$ depends on the bare parameters
(larger values of $\Lambda$ and $\lambda$ in this sense are preferable) and also 
on the accuracy $\Delta$ in the 'time' integration and on $\delta x$ in  
the 'spatial' mesh. However the process of improving 'space' and 'time' accuracy
has the effect of increasing the computer time  and in any case
the routine seems to be unable to converge for $k$ below some particular value.
So the best compromise found for the minimal value of $k$ in the present analysis 
is  $k=0.02$ at which the routine converges for all the examples considered.

\step

As a first step, we set the boundary values $\Lambda=10$ and $\lambda=0.1$
in Eq. (\ref{explbare}) and analyze the derivative of the potential
as obtained from the two coupled PDEs for $N=1$ (and $d=4$).
The results for $V^\prime_k(\phi)$ are shown in
Fig. \ref{uno}.  Curve {\bf (a)} is $V^\prime_\Lambda$, while
{\bf (b),\,(c),\,(d),\,(e)}
respectively correspond to $V^\prime_k$ at $k=0.7,\,0.5,\,0.3,\,0.02$.

\step

Together with these RG outputs we also display  in Fig. \ref{uno} the derivative
of the one-loop effective potential, {\bf (f)}, obtained with the same set
of renormalized  parameters of curve {\bf (e)}. More specifically,
the one loop quantum correction to the  effective potential,
properly regularized by a four-momentum cut-off $\Lambda$, reads (see e.g.\cite{jackiw})
\beq\label{v1loop}
V_{1l}(\phi)=\frac{1}{64\pi^2}\left \{ \Lambda^4\; {\rm ln}
\left (1+\frac{V^{\prime\prime}_o (\phi)}
{\Lambda^2} \right ) - \left (V^{\prime\prime}_o (\phi) \right )^2 \;
{\rm ln}  \left (1+\frac{\Lambda^2}{V^{\prime\prime}_o (\phi)}
 \right ) + \Lambda^2 V^{\prime\prime}_o(\phi)\right \}
\eeq
where the prime, again, denotes the derivative with respect to the field and
$V_o$ is the bare potential which, in our case, is given in
Eq. (\ref{explbare}). In the presence of a negative square mass
term and therefore of unstable modes, the logarithms in Eq. (\ref{v1loop})
develop an imaginary part and here  we shall only consider the real part of 
the one loop effective potential.

\step

We observe  that  the flow equation has a different dependence on the
the ultraviolet cut-off $\Lambda$ with respect to Eq (\ref{v1loop}).
This can be directly checked by integrating Eq. (\ref{vflow}) in the
LPA (i.e. fixing $Z_k=1$) and also neglecting the $k$ dependence of
$V_k$ in the right hand side of the equation, in order to preserve
at least the leading dependence on $\Lambda$. In this case the flow
equation reads:
\beq \label{simplerflow}
\partial_k V_k(\phi) =  \frac{k^5}{32 \pi^2}\frac{1}{k^2+V^{\prime\prime}_o(\phi)} \; ,
\eeq and its integration from $k=\Lambda$ down to $k=0$ gives the
following expression of  the correction to the first derivative of
the bare potential: \beq\label{correz}
V^{\prime}_{corr}(\phi)=\frac{V^{\prime\prime\prime}_o}{64\pi^2}
\left \{ \Lambda^2 -2 V^{\prime\prime}_o \; {\rm ln} \left
(1+\frac{\Lambda^2}{V^{\prime\prime}_o (\phi)} \right ) +
\frac{\Lambda^2 V^{\prime\prime}_o(\phi)}{\Lambda^2
+V^{\prime\prime}_o(\phi)} \right \} \eeq to be compared with the
derivative of Eq (\ref{v1loop}): \beq\label{correz1l}
V^{\prime}_{1l}(\phi)=\frac{V^{\prime\prime\prime}_o}{64\pi^2} \left
\{ 2 \Lambda^2 -2 V^{\prime\prime}_o \; {\rm ln} \left
(1+\frac{\Lambda^2}{V^{\prime\prime}_o (\phi)} \right ) \right \} \;
.\eeq 
While the logarithmic term in Eqs. (\ref{correz}) and
(\ref{correz1l}) is the same, the leading term proportional to
$\Lambda^2$ has different coefficients. Eq. (\ref{correz})  also
shows an additional subleading term. However, the difference between Eqs.
(\ref{correz}) and (\ref{correz1l}) plays no role for the
renormalized quantities because it can  be totally compensated by a
suitable choice of the counterterms so that the absence of
divergences (i.e. the cancellation of any dependence on the cut-off
$\Lambda$) ensures the smallness of the one loop corrections to the
classical potential. In practice this is realized by fixing  the proper 
mass and coupling counterterms in the one loop effective potential,
so that its second and fourth derivatives
computed at the minimum of the one loop potential, 
$\bar \phi_{1l}$, respectively coincide with
the numerical values of the (right limit) of the second and 
fourth derivative of the RG generated potential at $\bar\phi$,
as obtained from curve {\bf (e)}. 
As it is evident from Fig. \ref{uno}, curve {\bf (f)}
obtained by the above renormalization procedure,
is extremely close to curve {\bf (e)} in the region above the minimum, 
with the small differences  mainly referable to higher loop corrections 
and to $1/\Lambda^2$ terms that are neglected in our computation of the 
renormalized  one loop potential.
Instead, as is well known, the large difference between {\bf (e)} 
and {\bf (f)} in the region below the minimum is due to the
failure of the perturbative expansion in recovering the convexity
property of the effective potential which corresponds to the flat
region of curve {\bf (e)}.

\step

Before proceeding we also quote the perturbative correction of the field
renormalization as obtained from a one loop calculation in \cite{ilio},
\beq\label{z1loop}
Z_{1l}(\phi)=\frac{\lambda}{4\pi^2}\left \{ \frac
{V^{\prime\prime}_o (\phi)- V^{\prime\prime}_o (0)}
{V^{\prime\prime}_o (\phi)} \right \}
 \; ,
\eeq
which has to be added to the leading term  $Z_o=1$.
Eq. (\ref{z1loop}), when computed at the physical value of the field,
i.e. at the minimum of the potential $\bar \phi$, vanishes in the disordered
symmetric phase ( $\bar \phi=0$ ), but it has a finite value in the broken phase,
proportional to the coupling $\lambda$.
In particular, in our example with $\lambda=0.1$ , the one loop correction is
$Z_{1l}(\bar \phi)= 1.69\, 10^{-3}$.

\step

Let us come back to the RG generated potential. In the region of small $\phi$
the curves {\bf b,\,c,\,d,\,e} show the flattening which had been already
observed many times \cite{aoki,tetradis2,d1,bonannolac,Litim:2006nn,consolizap,caillol}:
the evolution of the potential is smooth
until $k$ reaches the infrared threshold where
$k^2\simeq - V_k^{\prime\prime}(\phi=0) > 0$, i.e. the region of the
unstable modes which induce strong modifications to the propagator and
therefore to the flow. When $k$ becomes smaller than this threshold,
$V^\prime_k(\phi)$ starts to develop a linear behavior in $\phi$ close to
$\phi=0$ (see curve {\bf b}) and, when $k$ gets smaller, the linear region
extends to larger values of $\phi$ with a slope that decreases toward zero.
On the other hand,
the region with $\phi>\bar\phi$ is substantially $k$ independent for 
$k$ below threshold, with only very small changes when $k\sim 0$
(see curves {\bf (c),\,(d),\,(e)}). The region between the two regimes at small
and large $\phi$ shows a sudden change of slope that becomes sharper and sharper 
for lower values of $k$. 
However when the details of this sudden change are properly enlarged,
the field derivatives of curves 
{\bf (b),\,(c),\,(d)}  
still show a continuous  behavior while
for  the derivative of curve {\bf (e)}, obtained for $k=0.02$,
the change is so sharp that, according to the numerical precision imposed,
the routine  is not able to approach zero beyond $k=0.02$. 

\step
 \begin{figure}
${}$\vskip1cm \epsfig{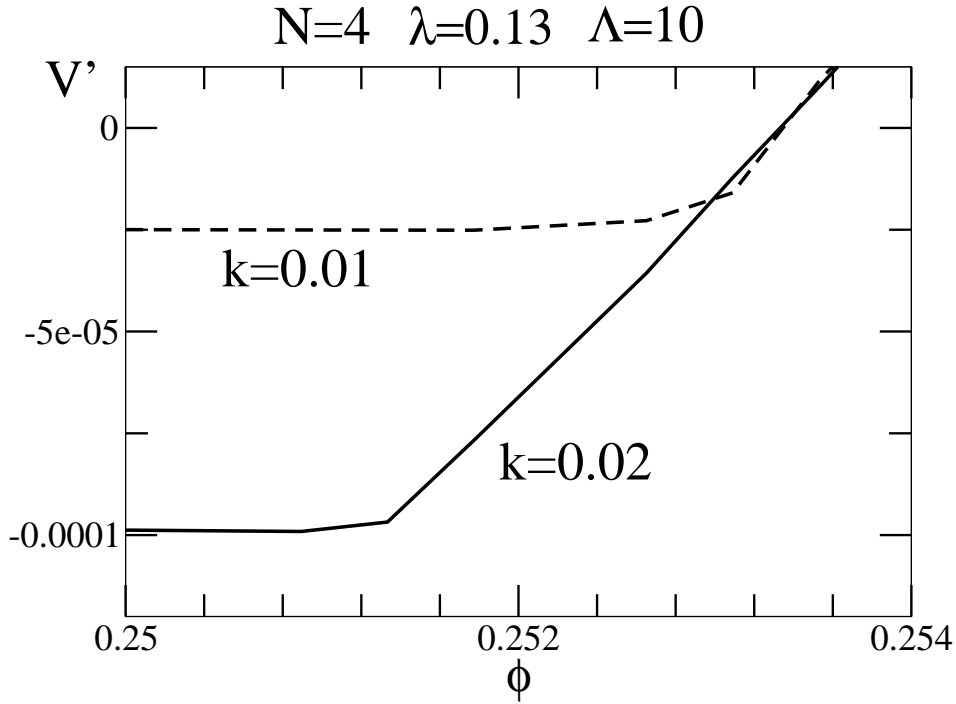}
\caption{\label{unobis}
$V^{\prime}_k$ for $N=4$ and $d=4$ at $k=0.02$ and $k=0.01$ in LPA  with  
$\lambda=0.13$, $\Lambda=10$.
}
\end{figure}

This is essentially what was expected.
Indeed, it is known that at $k=0$ the potential must reproduce the features 
of the effective potential which is convex, with a flat region
for  $- \bar \phi \leq \phi \leq\bar \phi$ and with a non-analytic  behavior,
that consists in a jump of the second derivative of the potential at $\bar \phi$
from zero to a finite value. As a consequence, the computation of
the zero momentum Green functions from the derivatives of the effective potential
is typically obtained by taking the limit of the derivatives from above,
$\phi \to \bar\phi$  with $\phi>\bar\phi$,
while the limit from below (with $\phi<\bar\phi$) is associated to
tunneling processes with infinitely large time scales.
In our numerical analysis the 'flat' part of curve ({\bf e}) obtained at $k=0.02$,
is practically vanishing: it differs from zero for less than $10^{-3}$.
Moreover, the fact that the positive branches of  {\bf (c),\,(d),\,(e)} 
are practically coincident, implies that these curves provide a reasonably good 
estimate of $\bar\phi$ even if the flow has not reached the point  $k=0$.

\step

We are also able to check $V_k(\phi)$ for small $\phi$ and $k$. In
fact, in this regime the relation $V^{\prime}_k(\phi)= - k^2 \phi$
has been repeatedly derived starting from the functional form of the
flow equations in the LPA 
(see e.g. \cite{ringwald,tetradis1,alexander,bonannolac,caillol}). 
In the example displayed in Fig. \ref{uno}, 
the effect of $Z_k$ is included and we have no analytical solution
for the coupled PDEs; however for the curves {\bf
(b),(c),(d)} and {\bf (e)} we find  : $- V^{\prime}_k(\phi)/ (k^2
\phi)=0.955(10)$ at $\phi=0.1$ and  $- V^{\prime}_k(\phi)/ (k^2
\phi)=0.930(10)$ at $\phi=0.5$ and this gives an estimate of the
size of the corrections to the above relation, which stay below
$10\%$ even for values of $k$ and $\phi$ close to $\bar \phi /2$.

 \begin{figure}
${}$\vskip1cm \epsfig{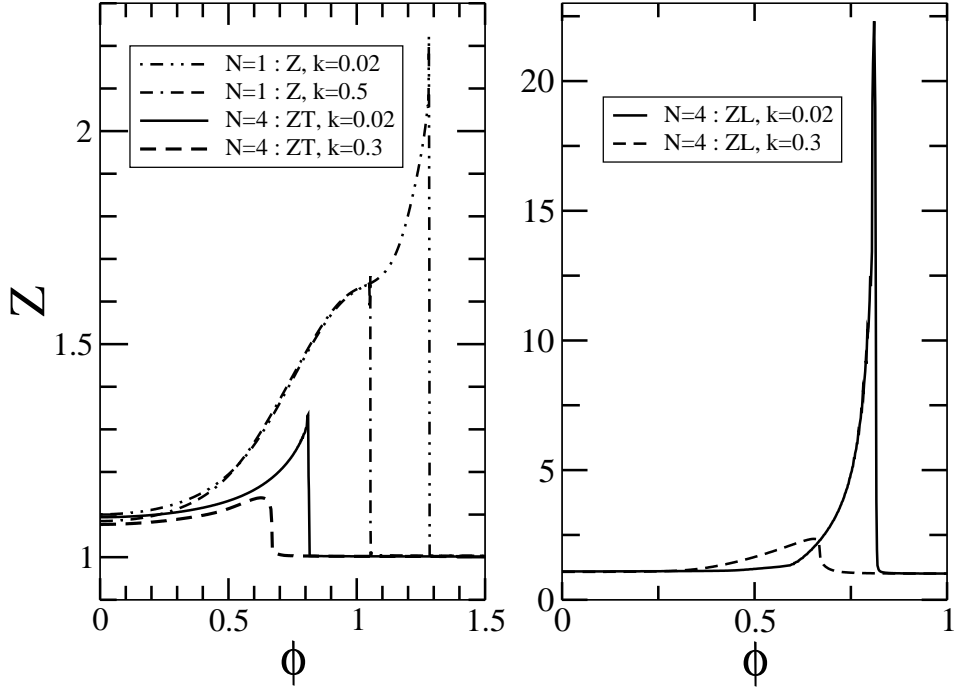}
\caption{\label{due} Left panel:
 $Z_k$ for $N=1$ at $k=0.5,\,0.02$ (dot-dashed);
$Z_T$ for $N=4$ at $k=0.3$ (dashed) and $k=0.02$ (continuous).
Right panel: 
$Z_L$ for $N=4$ at $k=0.3$ (dashed) and $k=0.02$,
(continuous).  $d=4$, $\Lambda=10$ and $\lambda=0.1$ for all curves.
\\[2ex]}
\end{figure}

\step

In the $N=4$ case the derivative of the potential shows a linear behavior 
in the range of small $\phi$ and $k$ similarly to the case $N=1$ in  Fig. \ref{uno}.
However for continuous symmetries in the ordered phase the presence of 
the Goldstone bosons has a strong impact on the infrared sector of the theory.
In particular, according to the study of the susceptibility of the $O(N)$-symmetric
theories ($N>1$) \cite{zinn,Anishetty:1995kj,Engels:1999dv}, 
in $d=4$ one expects a vanishing second derivative
of the potential at $\phi=\bar\phi$. This means that the non-analyticity developed in 
the  $N=1$ potential at $k=0$, discussed before, should disappear in  $N=4$.
To analyze this point, we plot in Fig. \ref{unobis} an enlarged detail of 
$V^{\prime}_k$  around $\bar\phi$, with  $N=4$, $d=4$, for $k=0.02$ and $k=0.01$. 
This computation is performed in the LPA  at $\lambda=0.13$ and $\Lambda=10$, 
in order to get closer
to $k=0$ and have a more accurate check on the behavior of $V^{\prime}_k$.   
In fact, in  Fig. \ref{unobis} the magnification is so large that the various mesh points
can be identified but, both  at $k=0.02$ and $0.01$,  $V^{\prime}_k$ shows a smooth 
growth without apparent jumps or rapid numerical fluctuations which were instead 
observed in the $N=1$ case. According to the numerical precision considered we
can state that the LPA  solution is approaching $V^{\prime\prime}_{k=0}(\bar\phi)=0$.
The inclusion of the field renormalization makes more difficult the approach to 
$k=0$ but there is no qualitative modification of the picture observed in the LPA
up to $k=0.02$.

\step

Let us now focus on the other variables, that is
the field renormalization $Z_k$ and $Z_L,\,Z_T$.
The qualitative  behavior of the field renormalization
is already  illustrated in \cite{consolizap}.
It is essentially perturbative when $k$ is larger than the mentioned threshold,
with deviations of order $10^{-3}$ from 1 (see Eq. (\ref{z1loop}) ).
Then, when $k$ decreases, a bump appears for values of $\phi$ 
around the separation point of the two regimes of $V_k^{\prime}$. 
When $k$ approaches zero this bump grows, dropping  very rapidly to the 
standard perturbative value at the separation point.
In Fig. \ref{due} the field renormalization in $N=1$ and $N=4$ (and $d=4$)
are shown. In the left panel of Fig. \ref{due}
we collected $Z_k(\phi)$ for $N=1$, at $k=0.5$ and $k=0.02$ together 
with  $Z_T$ for $N=4$, at $k=0.3$ and  $k=0.02$. In the right panel 
$Z_L$ for $N=4$, is plotted  at $k=0.3$ and  $k=0.02$ with a much larger 
scale on the $y$-axis because of the very high peak of $Z_L$.
 
\step

\begin{figure}
${}$\vskip1cm \epsfig{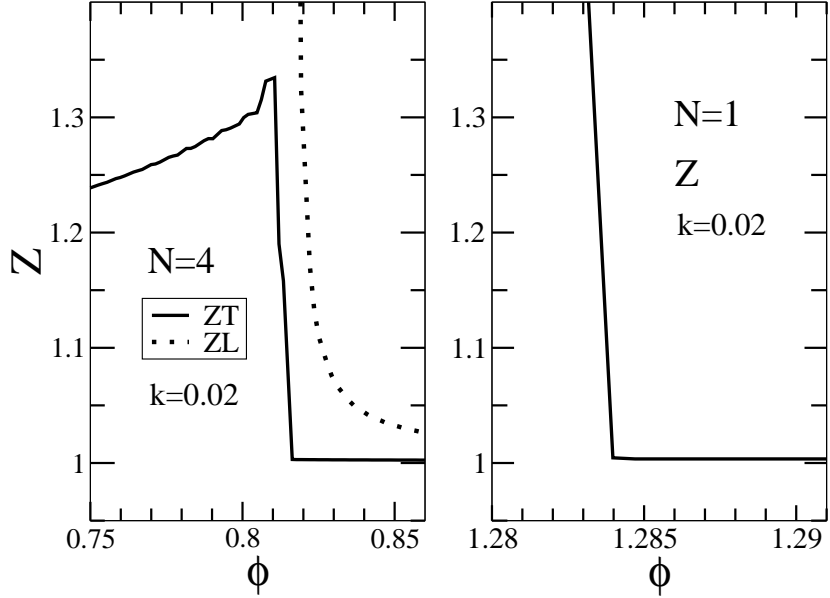}
\caption{\label{tre}
Enlarged details of Fig.~\ref{due}. Left panel:
$Z_T$ (continuous) and $Z_L$ (dotted) for $N=4$ at $k=0.02$.
Right panel:
$Z_k$ (continuous) for $N=1$ at $k=0.02$.
$d=4$, $\Lambda=10$ and $\lambda=0.1$ for all curves.
\\[2ex]}
\end{figure}

Before going on, it must be remarked that the peak  of
$Z_k(\bar \phi)$ for $N=1$ in Fig. \ref{due}  is much smaller
than that observed in \cite{consolizap} (whose value was around $10$), where
the proper time RG flow  with sharp cut-off on the proper time variable was used.
That particular version of flow equation, due to its  exponential nature,
although very rapidly converging and very accurate in computing
the critical exponents of the Ising universality class,
was criticized in \cite{bonannolac} for not well reproducing the details
of the singularity in the second derivative of the potential in the broken phase.
According to this criticism, the large size of the peak found
in \cite{consolizap} could be addressed to  that version of the flow equation
which artificially  amplifies the behavior of $Z_k$ close to $\bar\phi$.
On the other hand the approach followed  in this paper is expected to be
more accurate at quantitative level and therefore more suitable to describe
this particular feature.

\begin{figure}
${}$\vskip1cm \epsfig{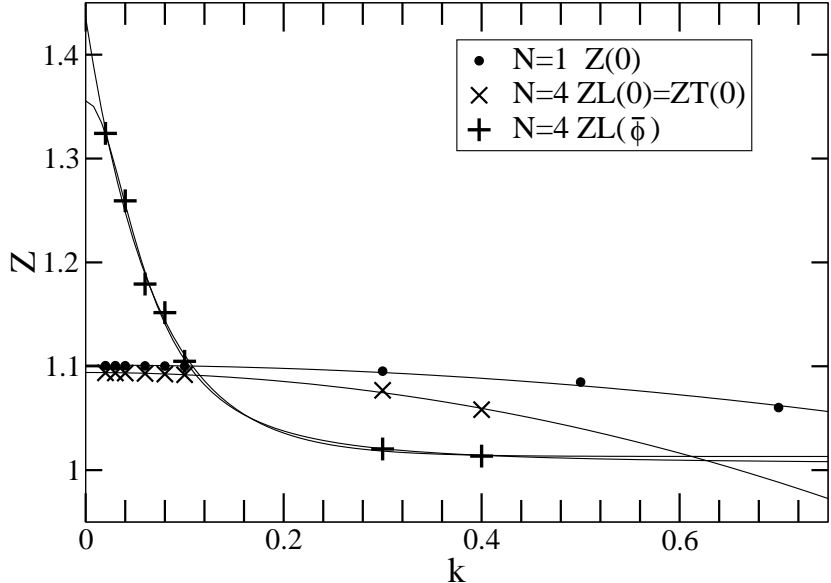}
\caption{\label{quattro}
$Z_k(\phi=0)$ for $N=1$ (black circles {\bf  $\bullet$})
and  $Z_L(\phi=0)=Z_T(\phi=0)$  for $N=4$ (crosses   {\bf \Large $ \times$})
and  $Z_L(\bar\phi)$ for $N=4$ ( plus  {\bf \large $+$})
at various values of $k$. In all cases $d=4$, $\Lambda=10$ and $\lambda=0.1$.
 Fits to these data are also plotted (see text).
\\[2ex]}
\end{figure}

\step

Going back to the curves of the field renormalization,
beside the size of the peaks other differences 
are illustrated in  Fig. \ref{tre}, where an enlargement of the 
lowest part of Fig. \ref{due} (only those curves at $k=0.02$ are plotted) is shown.
We note that both $Z_k$ of the $N=1$ theory and the transverse component $Z_T$ of
the  $N=4$ theory show a very sharp drop followed by a drastic change of slope.
However, by comparing  the scale of the $x$-axis in the left and right panel of 
Fig. \ref{tre}, it is evident that $Z_k$ with $N=1$ is much steeper than $Z_T$ with $N=4$
and, in particular, the drop observed in the $N=1$ case is very close 
(within the numerical accuracy employed) to a discontinuous jump. 
Instead, the longitudinal 
component $Z_L$ of the $N=4$ theory smoothly
approaches the  perturbative regime  for $\phi \sim \bar\phi$ and  this,
as shown below, yields  a  large (if compared with the perturbative correction),
finite prediction of $Z_L(\bar\phi)$.

\step

\begin{figure}
${}$\vskip1cm \epsfig{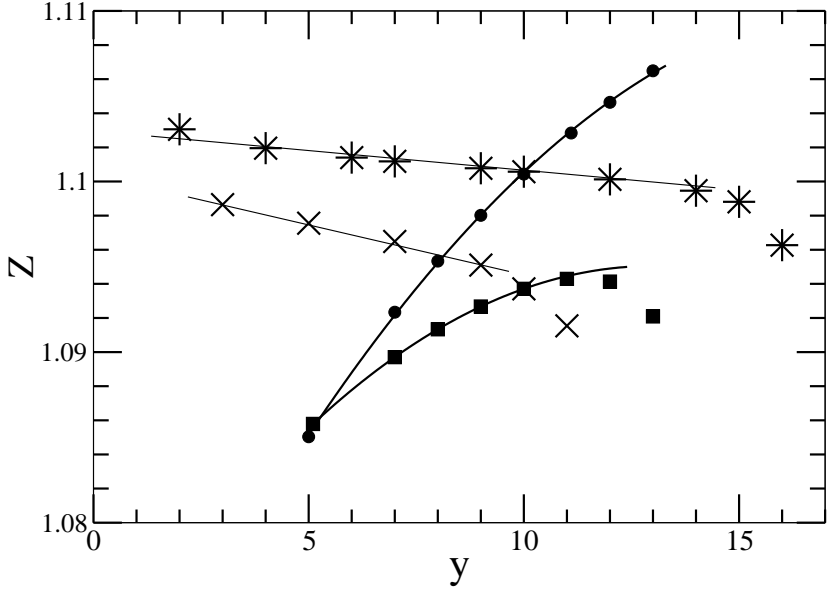}
\caption{\label{cinque}
$Z_k(\phi=0)$ for $N=1$ (black circles {\bf  $\bullet$})
and $Z_L(\phi=0)=Z_T(\phi=0)$ for $N=4$ (black squares $ \blacksquare$)
vs. $y=100\, \lambda$ at $k=0.02$ and $\Lambda=10$.
$Z_k(\phi=0)$ for $N=1$ (stars {\bf \Large  $*$})
and  $Z_L(\phi=0)=Z_T(\phi=0)$ for $N=4$ (crosses {\bf \Large $ \times$})
vs. $y=\Lambda$ at $k=0.02$ and $\lambda=0.1$. All curves are for $d=4$.
Fits to these data are also plotted (see text).
\\[2ex]}
\end{figure}

The structure of the flow equation of the field derivative is more complicated than
the equation of the potential  and  even an approximate analytic solution is missing.
However, from our numerical investigation we are now able to get some hints
on the dependence of $Z$  on various parameters such as  the running scale $k$
or the bare coupling  $\lambda$ or the UV cut-off $\Lambda$.
We concentrate on those values of $\phi$ where  $Z$
can be reasonably extracted, and therefore  we shall examine the
origin $\phi=0$ and the point $\bar \phi$.
We avoid instead  to analyze the peak of $Z$
which is often affected by large fluctuations.

\step

We start by observing that in our computation in Fig. \ref{tre},
for $N=1$  the minimum is at $\bar\phi=1.285$  and
one finds  $Z_k\sim 1.003$ for $\phi>1.284$ at $k=0.02$,
so that a typical perturbative value is observed at $\bar\phi$. 
Then one should conclude that  $Z_{k=0}(\bar\phi)\sim 1.003$,
unless at $k=0$ the sharp drop of $Z_k$ hits $\bar\phi$,
in which case no precise determination of $Z_k(\bar\phi)$ would be possible
within our numerical accuracy.
The data collected when $k$ approaches zero do not exclude either of these
two alternatives.
In the case of the $N=4$ theory one has  $\bar\phi=0.819$  and $Z_T\sim 1.003$
for $\phi>0.816$ at $k=0.02$ similar to what happens in the $N=1$ case
with the difference that for $N=1$ the jump could reasonably turn into a discontinuity,
as already noticed, while in $N=4$ the less steep drop reported in the left panel 
of Fig. \ref{tre} together with the smooth behavior of the potential around
$\bar\phi$ observed in Fig. \ref{unobis}, suggest that $Z_T$ could not develop 
a discontinuous gap at $k=0$.
On the other hand, as it is evident from the left panel of  Fig. \ref{tre},
the renormalization of the longitudinal field  is more regular and the large peak
decreases so smoothly  that at  $\bar\phi$ one finds $Z_L=1.324$ at $k=0.02$.
Then, in this case it is  possible to study the evolution of $Z_L(\bar\phi)$ as a
function of $k$ and eventually extrapolate its value at $k=0$.

\step

In Fig. \ref{quattro}  we show some values of $Z_k(\phi=0)$,
$Z_L(\phi=0)=Z_T(\phi=0)$, and of $Z_L(\bar\phi)$, obtained at small values of  $k$. 
In all cases $\lambda=0.1$, $\Lambda=10$ and $d=4$.
For the points at $\phi=0$  we found that  the functional form
\beq
f(k)=a\,k^2+b
\eeq
provides excellent fits to the data with $a=-0.079$, $b=1.101$ for $N=1$
and $a=-0.216$, $b=1.094$ for  $N=4$ and the corresponding plots are also shown
in Fig. \ref{quattro}.
As in the case of the potential, we observe here
a quadratic dependence on the running scale $k$.

\step

\begin{figure}
${}$\vskip1cm \epsfig{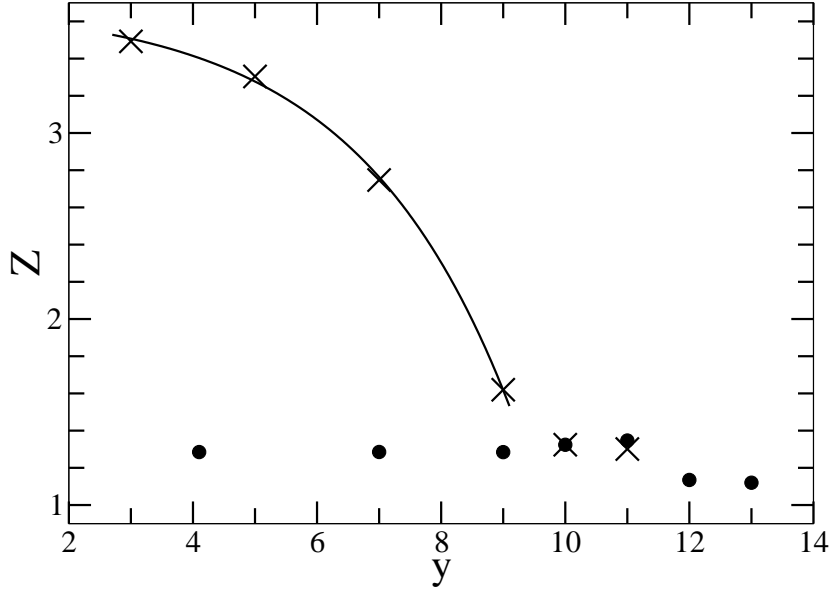}
\caption{\label{sei}
$Z_L(\bar \phi)$ for for $N=4$ (black circles  $\bullet$)
vs. $y=100\, \lambda$ at $k=0.02$ and $\Lambda=10$.
$Z_L(\bar \phi)$ for $N=4$ (crosses {\bf \Large $ \times$})
vs. $y=\Lambda$ at $k=0.02$ and $\lambda=0.1$. In all cases $d=4$.
A fit to the  latter set of data is also plotted (see text).
\\[2ex]}
\end{figure}

Let us now consider  $Z_L(\bar\phi)$.  In this case the two expressions
\beq
Z_L=1.0056 +\frac {0.0014}{k^2+0.0040} \;\;\;\;  \;\;\;\;  {\rm and} \;\;\;\;  \;\;\;\;
Z_L=1.0131 +0.4225 \; e^{-14.6466 \,k}
\eeq
fit equally well the data and we report both curves in Fig. \ref{quattro}.
Their extrapolation at $k=0$ yield two different (of about $7\%$) values of $Z_L$.
However, remarkably, the fitted points do not
suggest a divergent behavior of $Z_L(\bar\phi) $ at $k=0$, but, on the contrary,
indicate finite values not very distant from that found at $k=0.02$ .
We take this as a general indication that for this variable no strong modifications
should be expected in the range  $0< k< 0.02$.

\step

In Figure \ref{cinque} we show
$Z_k(\phi=0)$ and
$Z_L(\phi=0)=Z_T(\phi=0)$ versus $y=100 \,\lambda$ ,
at $k=0.02$ and $\Lambda=10$;
we also show  $Z_k(\phi=0)$ and
$Z_L(\phi=0)=Z_T(\phi=0)$ versus $y=\Lambda$
at $k=0.02$ and $\lambda=0.1$. Except for the black circles, the other points
show a rapid bend downward for large $y$ and this is due to the approaching of
the critical value of $\lambda$ or $\Lambda$, which signals the transition to the
symmetric phase. Apart from these points the two curves associated to circles
and squares have the form  $g(\lambda)=a \lambda^2+b\lambda+c$ which is a typical expansion
in powers of the coupling (for completeness we find
$a=-1.4823$, $b= 0.3855$, $c= 1.0700 $ for the squares
and $a=-1.4245$,  $b= 0.5214$, $c= 1.0626 $ for the circles),
while the points related to the field renormalization
dependence on $\Lambda$ display a linear behavior 
in a large range of the UV cut-off.

\step

Finally in Fig. \ref{sei} we show two  plots of  $Z_L(\bar\phi)$,
one versus $y=\Lambda$  with $\lambda=0.1$  and
the other versus $y=100 \,\lambda$  with $\Lambda=10$.
All points are collected at $k=0.02$.
We observe a small dependence on the coupling with a small increase
around $\lambda=0.1$ followed by a drop when approaching the critical line.
Instead, we find  a much stronger dependence on the cut-off with
the field renormalization that reaches $3.5$ for $\Lambda=3$.
By excluding the two points at large $\Lambda$, which are closer to the critical line,
the other points are in good agreement with the curve
$Z_L=3.694-0.0562 \, {\rm e}^{ 0.401 \, \Lambda}$,
which can obviously be expanded in a polynomial.
\step

\section{CONCLUSIONS}
\label{conclusions}

In this paper, we used the functional renormalization group
to analyze the renormalization function of the scalar field
in the $N=1$ theory and of the longitudinal and transverse components in the
$O(4)$-symmetric theory  in four Euclidean dimensions in the ordered phase where,
due to spontaneous symmetry breaking,
a non-vanishing vacuum expectation value  $\bar \phi\neq 0$ is generated.
In particular, the approximation scheme of
\cite {blaizot,Benitez:2009xg,Benitez:2011xx} on the RG
flow equations with the cut-off in Eq. ({\ref{optimctf}}) provides a
set of numerically stable coupled PDEs for $V$ and $Z$, which allow us to
approach the physical limit $k=0$ and to determine the main features of
the field renormalization.
\step

We found that a large non-perturbative enhancement of $Z$ occurs
in the range of $\phi$ corresponding to the observed flattening of the potential.
This effect in $Z$ had already been observed in quantum mechanics \cite{d1}
and its large value  in the classically forbidden region has also a direct
interpretation in terms of suppression factor of the tunneling probability of a
wave packet between two vacua \cite{curt,bra1}.
However, in field theory the numerical resolution
of the equation is more problematical and a first indication of a large field
renormalization for the $N=1$ case was obtained in \cite{consolizap}.
Our analysis qualitatively confirms those findings, but
the maximum value  of the peak  found here is  largely reduced with respect
to \cite{consolizap}.

\step

In particular we found that for $N=1$  the potential $V_k$ develops 
a non-analytic behavior at $\bar\phi$  in the limit $k\to 0$ 
and $Z_k$ presents a peak  
followed by  a sharp drop that occurs definitely for $\phi<\bar \phi$
at $k=0.02$ (which is the typical lower limit reachable by our numerical 
routine). If this should persist even at $k=0$, 
then  $Z_k(\bar \phi)$ would be substantially perturbative;
otherwise if at $k=0$ the drop  should occur at $\bar\phi$,
then we were not able to obtain a reasonable determination of 
 $Z_k(\bar \phi)$.
The data collected close to  $k\sim 0$ do not exclude either of these
two alternatives.

\step

In the $N=4$ theory the numerical analysis indicates that the potential does 
not develop a discontinuity in its second derivative with respect to $\phi$,
as in the  $N=1$ case, and even the transverse component $Z_T$ 
shows a less sharp plot than $Z_k$ for $N=1$, suggesting in this case a continuous 
behavior at $\bar\phi$ and $k\sim 0$. At the same time
the longitudinal component $Z_L$ has a very high peak but  the curve is so well behaved 
that it is possible to extract $Z_L(\bar\phi)$ at
various $k$ and its extrapolation down to $k=0$ remarkably leads to finite values.

\step

The dependence of $Z$ on the UV cut-off and on the bare coupling has also been investigated
and this provides  an indication on the typical numerical range spanned by the field
renormalization, although it must be recalled that, quantitatively,
these results are affected by large  uncertainties because even a very small numerical 
error in the estimate of $\bar\phi$ induce  large modifications in the determination
of $Z_L$.
\step


\begin{thebibliography}{10}

\bibitem{Polchinski:1983gv}
  J.~Polchinski,
  Nucl.\ Phys.\  {\bf B231 } (1984)  269-295.

\bibitem{Wetterich:1992yh}
  C.~Wetterich,
Phys.\ Lett.\  {\bf B 301} (1993) 90.

\bibitem{Wilson:1973jj}
  K.~G.~Wilson, J.~B.~Kogut,
  Phys.\ Rept.\  {\bf 12 } (1974)  75-200.


  \bibitem{Bagnuls:2000ae}
  C.~Bagnuls, C.~Bervillier,
  Phys.\ Rept.\  {\bf 348 } (2001)  91.
  [hep-th/0002034].


\bibitem{Berges:2000ew}
  J.~Berges, N.~Tetradis, C.~Wetterich,
  Phys.\ Rept.\ {\bf 363 } (2002)  223-386.
  [hep-ph/0005122].

\bibitem{Morris:1994ie}
  T.~R.~Morris,
  Phys.\ Lett.\ {\bf B 329}, (1994) 241
  [hep-ph/9403340].

\bibitem{Litim:2001dt}
  D.~F.~Litim,
  JHEP {\bf 0111} (2001) 059
  [hep-th/0111159].

\bibitem{blaizot}
  J.~-P.~Blaizot, R.~Mendez Galain, N.~Wschebor,
  Phys.\ Lett.\  {\bf B 632 } (2006)  571
[hep-th/0503103];
Phys. Rev.  {\bf E 74}, 051116 (2006); Ibid. {\bf E 74}, 051117 (2006).


\bibitem{Benitez:2009xg}
  F.~Benitez, J.~P.~Blaizot, H.~Chate, B.~Delamotte,
R.~Mendez-Galain and N.~Wschebor,
  Phys.\ Rev.\   {\bf E 80} (2009) 030103
[0901.0128 [cond-mat.stat-mech]].


\bibitem{Benitez:2011xx}
F.~Benitez, J.~P.~Blaizot, H.~Chate, B.~Delamotte,
R.~Mendez-Galain and N.~Wschebor,
Phys.\ Rev. {\bf E 85} (2012) 026707
[1110.2665 [cond-mat.stat-mech]].

\bibitem{Litim:2010tt}   D.~F.~Litim and D.~Zappal\`a,
Phys. Rev. {\bf D 83} (2011) 085009 [1009.1948 [hep-th]].


\bibitem{zinn} J. Zinn-Justin, {\em Quantum Field Theory and Critical
                 Phenomena}, Clarendon Press, Oxford, 1996;

\bibitem{Pelissetto}
  A.~Pelissetto and E.~Vicari,
  Nucl.\ Phys.\ B {\bf 522} (1998) 605
  [cond-mat/9801098];
%
  Nucl.\ Phys.\ B {\bf 540} (1999) 639
  [cond-mat/9805317];
%
%
  Phys.\ Rept.\  {\bf 368} (2002) 549
  [cond-mat/0012164].

\bibitem{ParisenToldin:2003hq}
  F.~Parisen Toldin, A.~Pelissetto and E.~Vicari,
  JHEP {\bf 0307} (2003) 029
  [hep-ph/0305264].


\bibitem{syma} K. Symanzik, Commun. Math. Phys. {\bf 16} (1970) 48.
\bibitem{ilio} J. Iliopoulos, C. Itzykson and A. Martin,
Rev. Mod. Phys., {\bf 47} (1975) 165.
\bibitem{curt} T.L. Curtright and C. B. Thorn, J. Math. Phys., {\bf 25} (1984) 541.
\bibitem{convexity} Y. Fujimoto, L. O'Raifeartaigh and G. Parravicini,
Nucl.Phys. {\bf B 212} (1983) 268; C.M. Bender and F. Cooper
Nucl.Phys. {\bf B 224} (1983) 403; D.J.E. Callaway, Phys. Rev. {\bf D 27} (1983)
2974; V. Branchina, P. Castorina and D. Zappal\`a, Phys. Rev. {\bf D 41} (1990)
1948.
\bibitem{ringwald}
A. Ringwald and C. Wetterich, Nucl.Phys. {\bf B 334} (1990) 506.
\bibitem{tetradis1}
N. Tetradis and C. Wetterich, Nucl.Phys. {\bf B 383} (1992) 197.
\bibitem{aoki}
K.-I. Aoki, A. Horikoshi, M. Taniguchi and H. Terao,
{\it Nonperturbative renormalization group and quantum tunnelling},
in : {Proceedings of the Workshop "The Exact
Renormalization Group"}, Faro, Portugal, Sept.  1998, (World
Scientific 1999), p. 194.  arXiv:hep-th/9812050.
\bibitem{alexander} J. Alexandre, V. Branchina and J. Polonyi, Phys.
Lett. {\bf B 445} (1999) 351.
\bibitem{tetradis2}
A. S. Kapoyannis and N. Tetradis, Phys. Lett. {\bf A 276} (2000) 225.

\bibitem{d1}
  D.~Zappal\`a,
  Phys.\ Lett. {\bf A 290}, (2001) 35
  [quant-ph/0108019].


\bibitem{bonannolac}
  A.~Bonanno and G.~Lacagnina,
  Nucl.\ Phys. {\bf B 693}, (2004) 36 
  [hep-th/0403176].


\bibitem{Litim:2006nn}
  D.~F.~Litim, J.~M.~Pawlowski and L.~Vergara,
{\it Convexity of the effective action from functional flows},
[hep-th/0602140].

\bibitem{consolizap}
  M.~Consoli and D.~Zappal\`a,
  Phys.\ Lett. {\bf B 641}  (2006) 368
  [hep-th/0606010].

\bibitem{caillol}
J.-M. Caillol,   Nucl.\ Phys. {\bf B 855} (2012) 854.

\bibitem{Litim:2000ci}
D.~F.~Litim,
Phys.\ Lett.\   {\bf B 486} (2000) 92, [hep-th/0005245].


\bibitem{Litim:2001up}
  D.~F.~Litim,
  Phys.\ Rev.\  {\bf D 64} (2001) 105007
  [hep-th/0103195].

\bibitem{Litim:2001fd}
  D.~F.~Litim,
  Int.\ J.\ Mod.\ Phys. {\bf A 16} (2001) 2081
  [hep-th/0104221].



\bibitem{Pawlowski:2005xe}
  J.~M.~Pawlowski,
  Annals Phys.   {\bf 322 } (2007)  2831.
  [hep-th/0512261].



\bibitem{Liao:1996fp}
S.~B.~Liao,
Phys.\ Rev.\  {\bf D 53} (1996) 2020,
[hep-th/9501124].


\bibitem{Bohr:2000gp}
  O.~Bohr, B.~J.~Schaefer, J.~Wambach,
  Int.\ J.\ Mod.\ Phys.\  {\bf A 16 } (2001)  3823.
  [hep-ph/0007098].

\bibitem{Bonanno:2000yp}
  A.~Bonanno and D.~Zappal\`a,
  Phys.\ Lett.    {\bf B 504} (2001) 181,
  [hep-th/0010095].

\bibitem{Litim:2001hk}
D.~F.~Litim and J.~M.~Pawlowski,
Phys.\ Lett.  {\bf  B 516} (2001) 197, [hep-th/0107020].



\bibitem{Mazza:2001bp}
M.~Mazza and D.~Zappal\`a,
Phys.\ Rev.  {\bf D 64} (2001) 105013, [hep-th/0106230].

\bibitem{Bonanno:2004sy}
  A.~Bonanno and M.~Reuter,
  JHEP {\bf 0502} (2005) 035
  [hep-th/0410191].


\bibitem{Litim:2001ky}
  D.~F.~Litim and J.~M.~Pawlowski,
  Phys.\ Rev.   {\bf D 65} (2002) 081701
  [hep-th/0111191].

\bibitem{Litim:2002xm}
  D.~F.~Litim and J.~M.~Pawlowski,
  Phys.\ Rev.   {\bf D 66} (2002) 025030
  [hep-th/0202188].

\bibitem{Litim:2002hj}
  D.~F.~Litim and J.~M.~Pawlowski,
  Phys.\ Lett.   {\bf B 546} (2002) 279  [hep-th/0208216].

\bibitem{d2}
  D.~Zappal\`a,
  Phys.\ Rev.  {\bf D 66} (2002) 105020  [hep-th/0202167].

\bibitem{nag} M.~Berzins and P.~M.~Dew, ACM Trans. Math. Software, {\bf 17} (1991) 178.

\bibitem{jackiw} R. Jackiw, Phys. Rev. {\bf D 9} (1974) 1686.

\bibitem{Anishetty:1995kj}
  R.~Anishetty, R.~Basu, N.~D.~Hari Dass and H.~S.~Sharatchandra,
  Int.\ J.\ Mod.\ Phys.\ A {\bf 14} (1999) 3467
  [hep-th/9502003].

\bibitem{Engels:1999dv}
  J.~Engels and T.~Mendes,
  Nucl.\ Phys. {\bf B 572} (2000) 289.
  [hep-lat/9911028].


\bibitem{bra1} A. Bonanno and D. Zappal\`a, Phys. Rev. {\bf D 57} (1998) 7383;
A. Bonanno, V. Branchina, H. Mohrbach and D. Zappal\`a,
Phys.\ Rev.\ D {\bf 60} (1999) 065009
[hep-th/9903173];
G. Andronico, V. Branchina and D. Zappal\`a, Phys. Rev. Lett.
{\bf 88} (2002) 178902 [quant-ph/0205067];
V. Branchina, H. Faivre and D. Zappal\`a,
Eur. Phys. J. {\bf C 36} (2004) 271 [hep-th/0306050].

\end{thebibliography}
\end{document}